# A High-Performance Mid-infrared Optical Switch Enabled by Bulk Dirac Fermions in $Cd_3As_2$

Chunhui Zhu[1#], Fengqiu Wang[1#*], Yafei Meng[1#], Xiang Yuan[2], Faxian Xiu[2*], Hongyu Luo[3], Yazhou Wang[3], Jianfeng Li[3*], Xinjie Lv[4], Liang He[1], Yongbing Xu[1], Yi Shi[1], Rong Zhang[1], and Shining Zhu[4]

[1] School of Electronic Science and Engineering and Collaborative Innovation Center of Advanced Microstructures, Nanjing University, Nanjing 210093, China

[2] State Key Laboratory of Surface Physics and Department of Physics, Collaborative Innovation Center of Advanced Microstructures, Fudan University, Shanghai 200433, China

[3] School of Optoelectronic Information, University of Electronic Science and Technology of China, Chengdu 610054, China

[4] National Laboratory of Solid State Microstructures and School of Physics, Nanjing University, Nanjing 210093, China

[#] These authors contributed equally to this work.

*Correspondence Emails: fwang@nju.edu.cn; faxian@fudan.edu.cn; lijianfeng@uestc.edu.cn

**Pulsed lasers operating in the 2-5 μm band are important for a wide range of applications in sensing, spectroscopy, imaging and communications[1,2]. Despite recent advances with mid-infrared gain media, the lack of a capable pulse generation mechanism, *i.e.* a passive optical switch, remains a significant technological challenge. Here we show that mid-infrared optical response of Dirac states in crystalline $Cd_3As_2$, a three-dimensional topological Dirac semimetal (TDS)[3-5], constitutes an ideal ultrafast optical switching mechanism for the 2-5 μm range. Significantly, fundamental aspects of the photocarrier processes, such as relaxation time scales, are found to be flexibly controlled through element doping, a feature crucial for the development of convenient mid-infrared ultrafast sources. Although various exotic physical phenomena have been uncovered in three-dimensional TDS systems[6-9], our findings show for the first time that this emerging class of quantum materials can be harnessed to fill a long known gap in the field of photonics.**

Short-pulsed lasers have proved indispensable for many branches of science and engineering. The key component to achieve pulsed operation is a passive optical switch, also termed saturable absorber, which can transit between different absorption states on an ultrafast time scale[10]. While an optical switch may take different physical forms, Semiconductor Saturable Absorber Mirrors (SESAMs), a breakthrough in ultrafast photonics in the early 1990s[11], are at present the most prevalent approach used for ultrashort pulse generation in the near-infrared. Compared with alternative technologies, a most compelling advantage of SESAMs is the ease with which device parameters can be precisely customised with great reproducibility[12], thanks to the use of highly mature semiconductor growth techniques, such as Molecular Beam Epitaxy (MBE). Equally important, the design freedoms of SESAMs have opened up a desirably large parameter space, enabling the access to robust pulsation regimes and continuous improvement of output characteristics for near-infrared ultrafast lasers[10].

The development of compact short-pulsed lasers in the mid-infrared has historically been hindered by the poor availability of gain materials, thus expedient techniques based on near-infrared sources and nonlinear frequency conversion become today's norm for mid-infrared pulse generation. The rapid maturing of mid-infrared gain platforms in recent years is calling for saturable absorbers with performance levels on a par with their near-infrared counterparts. Unfortunately, due to the use of III-V compound semiconductors, SESAMs of conventional design operate only within the near-infrared, typically below 1.5 μm[10]. Although using quantum-dots structures[13] and narrow bandgap materials can extend the spectral coverage to ~ 2-3 μm[14], strategies for scaling the operation wavelength further into the mid-wave infrared remain elusive. Low-dimensional materials, including carbon nanotubes[15], graphene[16,17], transition-metal dichalcogenides[18], and other emerging two-dimensional materials have been considered for low-cost substitutes for SESAMs. However, the defect-prone exfoliation and transfer processes inevitably lead to poor repeatability and reliability, limiting the prospect for practical exploitation. Thus, a robust and broadband material proposal that can be generalised for the mid-infrared is still missing from the toolbox for implementation of compact mid-infrared pulsed lasers, and is of vital importance for enabling ultrafast photonic capabilities beyond the conventional near-infrared range.

$Cd_3As_2$, a representative three-dimensional Dirac semimetal that is deemed bulk analogue of graphene exhibits stable bulk Dirac states at room temperature[4,5]. Very recently, optical conductivity studies show that photoexcitations in the 1.25-5 μm correspond to interband transitions between the two Dirac bands in $Cd_3As_2$[19], which promises a highly robust and amendable light-Dirac fermion-interaction platform. Here, by probing the mid-infrared optical response of bulk Dirac fermions, we show that MBE-grown $Cd_3As_2$ can act as an ultrafast (< 10 ps) optical switch with operation covering at least the 2-5 μm range. Through element doping, a desirably large tuning of photocarrier recovery time was achieved, *i.e.* from 8 ps to 800 fs at 4.5 μm. Furthermore, the on-demand parameter customisation of $Cd_3As_2$ allows the access to different pulsation regimes in a 3 μm fluoride fibre laser, demonstrating the potential of significantly upgraded levels of optimization for mid-infrared pulsed lasers.

We first prepared high quality $Cd_3As_2$ films under ultra-high vacuum in MBE system (see Methods). The thickness of the film was *in-situ* monitored by reflection high-energy electron diffraction (RHEED). As shown in Fig. 1a inset, a typical RHEED pattern of $Cd_3As_2$ films shows bright and streaky lines, indicating a good surface morphology and crystallinity. X-Ray diffraction (XRD) measurements were also performed (Fig. 1a). A series of peaks can be well resolved and indexed as {112} crystal plane (the un-indexed peaks come from the mica substrate), further confirming the good crystallinity of the sample[20].

To confirm the existence of Dirac fermions, $Cd_3As_2$ thin films were subsequently patterned into standard six-probe Hall bar geometry, and the transport measurements were conducted using physical property measurement system with a superconducting magnet (9 T). Magnetoresistance (MR) is plotted versus magnetic field ($B$) in Fig. 1b. At low temperature, the MR ratio reaches 130 % within 9 T. A meticulous study of the MR curves reveals Shubnikov-de Haas (SdH) oscillations (Fig. S1a), the peak of which are denoted as the integer Landau index as shown in Fig. 1c. All these data points fall on a straight line. By performing a linear fit, the intercept is derived as $-0.04\pm0.06$, corresponding to the non-trivial Berry phase. This is a distinct feature of ultra-relativistic Dirac fermions which is shared by topological insulators[21], graphene[22] and single crystal bulk $Cd_3As_2$[23]. By further performing the SdH oscillations at different temperatures, the cyclotron mass of the system can be estimated based on the formula $\Delta R(T) = \Delta R(0)/\sinh(\lambda(T))$, where the thermal factor $\lambda(T)$ is given by $\lambda(T) = 2\pi^2 k_B T m^*/\hbar eB$ and $k_B$ and $m^*$ are Boltzmann's constant and cyclotron mass, respectively. Best fitting to the data in Fig. 1d gives an effective mass of $m^*=0.021m_e$, where $m_e$ is the free electron mass. The vanishingly small effective mass further confirms the Dirac nature of $Cd_3As_2$ thin films. Moreover, angle-dependent SdH oscillations (Fig. S1b) show a quantum scattering time of ~95 fs, and Hall measurement revealed an electron mobility of ~3300 $cm^2V^{-1}s^{-1}$ at room temperature, and 5400 $cm^2V^{-1}s^{-1}$ at low temperature (2.5 K).

Photoexcitation and carrier relaxation are fundamental processes that govern the optical response of materials. For $Cd_3As_2$, an optical conductivity over broad mid-infrared wavelengths that is linked to the intrinsic Dirac band dispersion provides an important prerequisite for robust and tuneable light-matter interactions[19]. This is more desirable than Dirac fermions formed in the two-dimensional limit, which are typically susceptible to extrinsic perturbations or easily overwhelmed by the bulk conductance states[24]. Indeed, a flat reflectance band across 2-5 μm is obtained for our sample, confirming inter-Dirac bands photoexcitation[19] (Fig. S2 and Supplementary note S1). To investigate the optical switching characteristics of $Cd_3As_2$, mid-infrared ultrafast pump-probe spectroscopy is performed (see Methods). It should be noted that while degenerate probing is known to better approximate saturable absorber dynamics, non-degenerate measurements offer the advantage of convenient scaling of probe wavelengths, due to the ease with beam alignment. Using a non-degenerate configuration allows the probe wavelength to be reliably extended to 4.7 µm, the cutoff for the PbSe photodetector used in our setup. Fig. 2a and Fig. S3a show the non-degenerate and degenerate transient absorption curves, respectively (a 400 nm thick $Cd_3As_2$ film is used to avoid quantum confinement effect). Both measurements reveal photobleaching signatures arising from the same band filling effect[25], indicating that $Cd_3As_2$ exhibits saturable absorption over the 1.6-4.7 µm range. Pure mica substrate is found to yield no transient response under the same condition. Fig.2b displays the transient absorption probed at 2.5 µm, illustrating the correlation between the non-degenerate and degenerate photocarrier processes, where $\tau_{nondeg}$ is seen to reproduce the value of $\tau_{deg2}$. We therefore attribute the fast component $\tau_{deg1}$ to coherent nonlinear optical response resulting from carrier-carrier scattering[26], and the slow component $\tau_{deg2}$ ($\tau_{nondeg}$) to the incoherent response, typically associated with carrier-phonon interaction[27]. In addition, the ratio of the slow component ($\Delta T/T_0$ at 3 ps) to the fast component ($\Delta T/T_0$ at 0 ps) as a function of probe wavelength is also investigated. The increasing trend shown in Fig.S3b indicates that incoherent carrier processes play an increasingly dominant role as the excitation photon energy shifts closer to the Dirac node. Fig. 2c summarizes the fitted time constants, where $\tau_{nondeg}$ is seen to slow down (from 4 ps to 8 ps) as the probe wavelength increases in the mid-infrared. The elongation of recovery time with increasing probe wavelength is a typical feature characterizing Dirac fermion relaxations in graphene that can be well interpreted by microscopic modelling based on density-matrix formulism[28]. Probing further into the mid-infrared allows quantitative understanding of the microscopic carrier-phonon interactions in the TDS systems[19], but is beyond the scope of the current study. Nevertheless, to the best of our knowledge, this is the first report of a MBE-grown material that demonstrates ultrafast saturable absorption (< 10ps) across the entire 2-5 µm range.

In addition to broadband operation, flexible parameter customization is the single most important feature that differentiates SESAMs from other saturable absorber technologies and makes SESAMs adaptable to various laser formats, *i.e.* fibre, solid-state, or semiconductor chip lasers[10-12]. However, flexible parameter tuning has not been experimentally achieved for mid-infrared saturable absorbers reported so far. Among various device parameters, recovery time represents the most intrinsic property to the absorber material. Other properties, such as modulation depth, saturation intensity, and non-saturable absorption can typically be controlled by engineering either the recovery time or the device structure[12]. It should be noted that for saturable absorber operation, the slow component actually plays a more important role than the fast component, especially during the initial pulse formation stage[10]. Therefore, various strategies targeting at the tuning of the slow component of the relaxation time of SESAMs have been proposed[29-31], of which low-temperature growth[30] and post-growth ion-implantation[31] have proved effective. Element doping has recently been used to achieve topological-phase transition in topological insulators[32]. The altering of electronic properties of topological systems is expected to lead to strong impact on the optical behaviours. Here we introduce Chromium (Cr) as a dopant to the $Cd_3As_2$ film (see Methods). Fig. 2d presents the non-degenerate pump-probe results at 4.5 μm for $Cd_2As_3$ films for different Cr concentrations (up to 2%). Owing to extra trapping centres introduced by the Cr dopants, the photocarrier recovery times of Cr-doped $Cd_3As_2$ become appreciably faster at higher Cr concentrations, for all wavelengths within 2-5 μm (Fig.S4). The time constants as a function of Cr concentration are summarized in Fig. 2e. Without any particular optimization, a relaxation time tuning across an order of magnitude (*i.e.* from 8 ps to 800 fs at ~4.5 μm) is already achieved, which opens up the long sought-after parameter space for mid-infrared optical switches.

To demonstrate the effectiveness of the mid-infrared optical switching properties of $Cd_3As_2$, we set up a $Ho^{3+}/Pr^{3+}$ co-doped fluoride fibre laser operating around 3 μm (Fig. 3a, see Methods). First, the un-doped $Cd_3As_2$ film with relatively long time constant (~7 ps at a wavelength of ~3μm) was used. When the pump power reached 58 mW, continuous wave (CW) emission turned into Q-switched mode-locking (QSML) envelope as shown in Fig. 3b. With further increasing of the pump power, both the duration and period of the Q-switched envelope decreased as expected for Q-switching operation[33]. The QSML state changed swiftly into CW mode-locking at a pump power of 80 mW, and can be maintained up to a pump power of 290 mW as shown in Fig. 3c. The pulse period of 70 ns matched well with the calculated cavity round trip time, indicating that the laser was operating in the single pulse per round-trip condition. The optical spectrum of the mode-locked pulses was shown in Fig. 3d. A centre wavelength of 2860 nm and a full-width-half-maximum (FWHM) of 6.2 nm were achieved. Furthermore, the radio-frequency (RF) spectra of the pulses were also measured, confirming robust and stable mode-locking operation (Fig. 3e). The pulse duration was measured by a home-built mid-infrared autocorrelator and a pulse width of 6.3 ps is inferred (Fig. 3f). It is clear that stable mode-locking is achieved with the un-doped sample. Then the Cr-doped $Cd_3As_2$ films with shorter relaxation times were introduced into the cavity in turn. It is found that the threshold to achieve CW mode-locking increased with shortening relaxation time of the $Cd_2As_3$, as higher intensities are now required to saturate the conduction band[12]. More specifically, in the case where the $Cd_2As_3$ film with the relaxation time of 0.5 ps was used (2% Cr doping concentration), only Q-switched pulsation regime was accessible as shown in Fig. S5. The ease with which to custom made the mid-infrared optical switch is expected to greatly facilitate more thorough studies of pulsation regimes of various mid-infrared lasers.

In conclusion, we have demonstrated a capable mid-infrared optical switch based on the emerging three-dimensional Dirac semimetal $Cd_3As_2$. Thanks to the unique optical response of bulk Dirac fermions and the compatibility with MBE growth, the use of $Cd_3As_2$ ensures broadband operation, flexible parameter control and synthesis scalability. These features effectively make $Cd_3As_2$ based approach a capable mid-infrared counterpart to the highly adaptable near-infrared SESAMs. In addition, the on-demand parameter customization further enables the mapping of absorber parameters to different pulsation regimes in a 3 μm fluoride fibre laser. Our work represents a significant step forward in the development of disruptive and convenient mid-infrared ultrafast sources for advanced sensing, communication, spectroscopy and medical diagnostics and may be extended to more sophisticated active photonic devices including mid-infrared optical modulators and light emission devices.

**Methods**

**$Cd_3As_2$ films growth:** The $Cd_3As_2$ films were grown under ultra-high vacuum in MBE system. Mica substrates were freshly cleaved before degassing at 350 ℃. The substrate temperature was set at 225 ℃ and epitaxial growth was carried out using co-evaporating of Cd (99.9999%) and As (99.9999%) from dual-filament and valve-cracker effusion cells. For Cr-doped sample, Cr source (99.999%) was introduced and the doping concentration is calibrated by energy-dispersion X-ray spectroscopy equipped in scanning electron microscope.

**Pump-probe measurement:** Both degenerate and non-degenerate pump-probe setup is based on a 1 kHz Ti: Sapphire amplifier system (Libra, Coherent Inc.). For the non-degenerate pump-probe experiment, a portion of the laser output energy is used to excite photocarriers in the sample, and the remaining part is fed into an optical parametric amplifier (OPA-SOLO, Coherent Inc.) to generate probe beam from 1.6 to 4.7 μm. For degenerate pump-probe measurement, the idler beam (1.6-2.6 μm) of the OPA is split into pump and probe. Both pump and probe pulses have durations of ~100 fs. The pump-induced change of probe is detected using a PbSe photodetector and a lock-in amplifier referenced to a 500 Hz chopped pump. The pump fluence was fixed at 100 μJ/cm$^2$. For non-degenerate setup, this intensity is in the saturation regime at a wavelength of 4.5 μm, but fall within the linear regime of 1.6 μm photoexcitation.

**3 μm fibre laser setup:** Two commercially available diode lasers (Eagleyard Photonics, Berlin) centered at ~1150 nm were employed to pump the gain fibre after polarization multiplexing via a polarized beam splitter (PBS) and then focusing by an AR-coated (@~1150 nm) ZnSe objective lens (Innovation Photonics, LFO-5-6-3 μm, 0.25 NA) with a 6.0 mm focal length. Note that this objective lens also acts as the collimator of the light out-coupled from the fibre core. A dichroic mirror (DM) with ~96% transmittance at 1150 nm and 95% reflectance at ~3 μm was placed between the PBS and ZnSe objective lens at an angle of 45° with respect to the pump beam to direct the laser. Another specifically designed dichroic mirror with 80% reflection at ~3 μm was used to act as the output coupler. A 3 μm filter with a FWHM of 500 nm was utilized to block the residual pump. The gain fibre (Fiberlabs, Japan) was a piece of commercial double-cladding Ho$^{3+}$/Pr$^{3+}$ co-doped fluoride fibre. It has an octangular pump core with a diameter of 125 μm and NA of 0.5 and a circular core with a diameter of 10 μm and NA of 0.2. The concentration of the Ho$^{3+}$ and Pr$^{3+}$ were 30,000 and 2500 ppm, respectively thus the selected fibre length of 6.8 m provided >90% pump absorption efficiency. Both ends of the fibre were cleaved at an angle of 8° to avoid parasitic lasing in the cavity. First, the laser from the angle-cleaved fibre end which was far from the pump source was collimated employing a specifically coating designed ZnSe objective lens (Innovation Photonics, LFO-5-12-3 μm, 0.13 NA) with a focal length of 12 mm (>95% transmission at 3 μm, and <10% transmission at 1150 nm). Then the collimated light was focused via another ZnSe objective lens the same as before onto the terminated feedback which was assembled by pasting the $Cd_2As_3$ films on a commercial gold-protected mirror (Thorlabs) as shown in the inset of Fig. 3a. Here the terminator was mounted onto a high-precision six-dimension adjuster to perform position optimization.

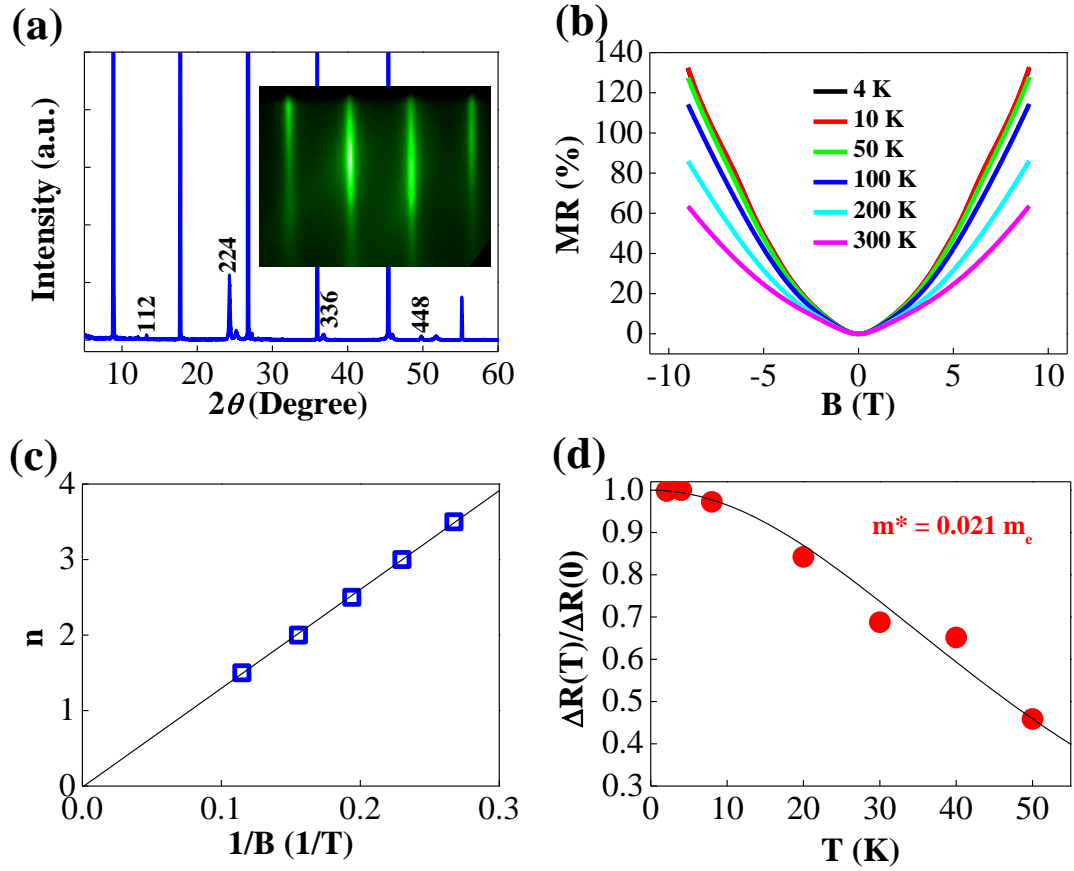

**Fig. 1 | Typical characterizations and magneto-transport of Cd$_3$As$_2$ thin films. (a)** X-ray diffraction pattern with {112} crystal surface. The inset is a typical in-situ RHEED pattern. **(b)** MR at different temperatures under perpendicular magnetic field. **(c)** A Landau fan diagram, showing non-trivial Berry phase. **(d)** Temperature-dependent oscillation amplitude.

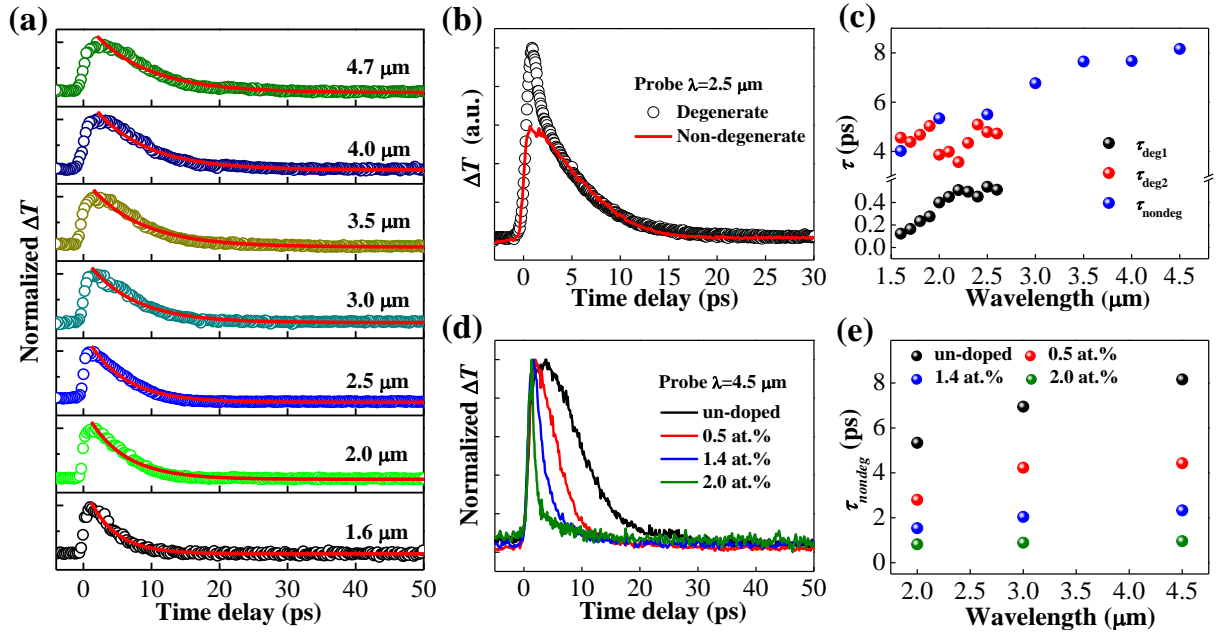

**Fig. 2 | Ultrafast nonlinear optical properties of Cd$_3$As$_2$ films. (a)** The non-degenerate ultrafast pump-probe results with the probe wavelength varying from 1.6 to 4.7 μm, and red solid lines correspond to a mono-exponential fit. **(b)** Time-resolved differential transmission spectra at 2.5 μm, showing the correlation between degenerate and non-degenerate relaxation processes. **(c)** Fitted relaxation time constants $\tau$ versus probing wavelengths for both the degenerate (black and red balls) and non-degenerate (blue balls) measurements. Note that, at relatively short wavelength, *i.e.* ~ 1.6 μm, the faster component of degenerate measurements $\tau_{deg1}$ is of the order of the pulse duration. Hence it may be not accurately resolvable. **(d)** Time-resolved $\Delta T$ traces at a probe wavelength of 4.5 μm for the Cd$_3$As$_2$ samples with different Cr concentrations, showing faster photocarrier relaxation time with higher Cr concentrations. **(e)** The fitted recovery time constants as a function of Cr concentration at 2.0, 3.0 and 4.5 μm.

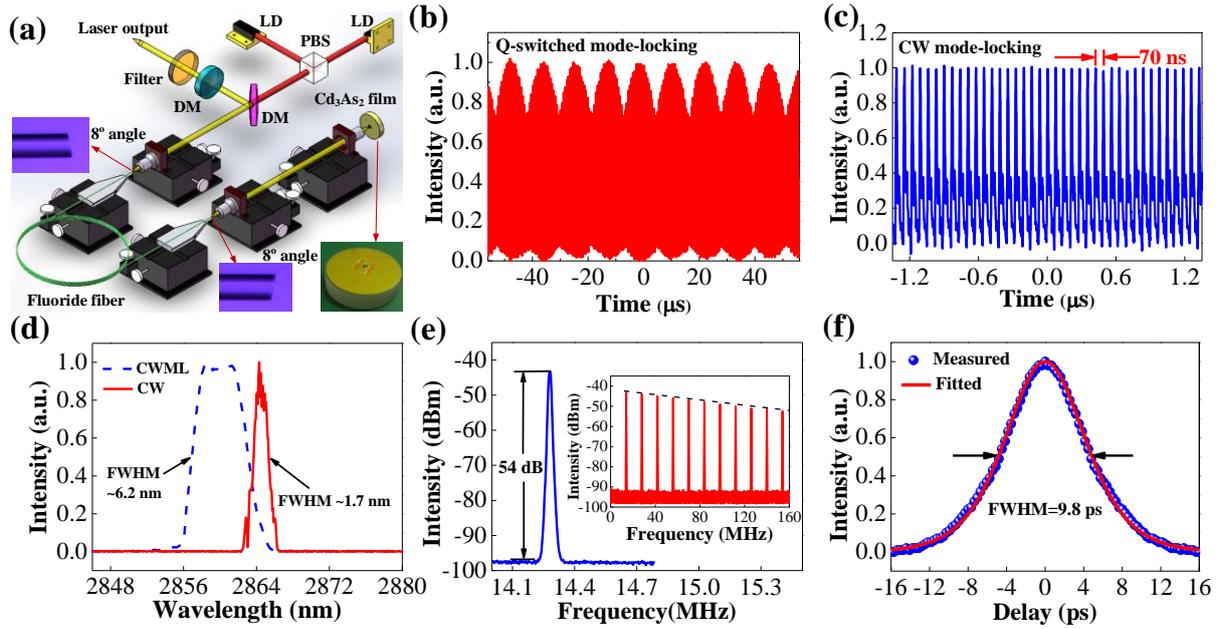

**Fig. 3 | Ultrafast mid-infrared fibre laser based on Cd$_3$As$_2$ saturable absorber. (a)** Schematic of the fibre laser setup, See method. **(b)** Q-switched mode-locked pulses at the launched pump power of 57.6 mW. **(c)** Continuous wave mode-locked pulses at the launched pump power of 286.9 mW. There is no discernable envelope modulation, indicating stable operation. **(d)** Output optical spectrum. It is observed that the centre wavelength of CW operation (by moving the focused beam onto a spot of the gold mirror clear of the Cd$_3$As$_2$ sample) was red-shifted to 2864.3 nm, as the decreased intra-cavity loss led to lower initial Stark manifold of the $^5I_6$ energy level. Meanwhile, the FWHM was decreased to 1.7 nm as a result of less Fourier spectral components required. **(e)** RF spectrum at a scanning span of 0.8 MHz with a resolution bandwidth of 10 kHz. The repetition rate and signal-to-noise ratio (SNR) were 14.28 MHz and 54 dB, respectively. The inset shows the RF spectrum with a broader scanning span ranging from 0 MHz to 160 MHz. The smooth roll-off of the clean harmonic frequency components indicated that no Q-switched modulation and multiple pulsing were presented in this operation regime. **(f)** Autocorrelation trace measured by intensity autocorrelator. The blue points are the experimental results and the red line is the fitting result using

*Sech*$^2$ function. The FWHM of the autocorrelation trace is 9.8 ps, corresponding to a pulse duration of 6.3 ps (a deconvolution factor of 1.54 is used to account for the *Sech*$^2$ pulse shape).